\documentclass{article}

\usepackage{epsf}

\begin{document}

\begin{center}

{\Large \bf 
Comparative statistical analysis of bacteria genomes in "word" context}

\vspace{0.5cm}

Olga V. Kirillova
\vspace{0.2cm}

Department of Theoretical Physics, St.Petersburg State University

Ulyanovskaya str. 1, St.Petersburg, 198904 Russia

(e-mail:kirill@heps.phys.spbu.ru)

\end{center}

\vspace{1cm}

{\small Abstract.     
Statistical analysis of bacteria genomes texts has been performed on
the basis of 20 complete genomes origin from Genebank.
It has been revealed that the word ranked distributions are quite well
approximated by logarithmic law. Results obtained in the absent words
investigation show the considerably nonrandom character of DNA texts. 
In character of autocorrelation function behavior in several
genomes period 3 oscillations were found. Short range autocorrelations
are present in short ($n=3$) words and practically absent in longer words.
}
\vspace{0.5cm}
    
PACS number(s): 87.10+e



\section{Introduction}
DNA molecules are main storage of information about any organism.
They are long sequences (linear or closed to a loop) contained in each
cell of an organism. Usually DNA sequences are represented by a string of
just four letters (A, C, G, T), each of them corresponds a definite type
of nucleotides: adenine, cytosine, guanine and thymine. These letters can
form different combinations. Purposely, some combinations in DNA texts are
nonrandom. They reflect structure and function of DNA and proteins. Where
from the question arises, what are regularities of such letter sequences
corresponded to known DNA properties?

Due to modern automatic techniques and new technologies of genome
sequencing one can observe
great increase of DNA texts data \cite{GB}. The crucial question of modern
genomics is what kind of information can be extracted from these data? In
this
realm many statistical methods were applied or even elaborated for
DNA sequences analysis. Great success was achieved
in DNA sequences classification \cite{JW},
autocorrelation analysis \cite{autocor,MA,Bor},
Fourier spectrum analysis \cite{MA, Voss, Bu1, Bu2},
wawelet analysis \cite{wawlet},
entropies calculations \cite{H94,Rowe,Gl},
methods of Hurst index estimation \cite{Lu,Yu}, transition matrix
analysis \cite{Yu}, random walk \cite{Bu2,rw},
usage of the mutual information function \cite{Her97,CSF},
detrended fluctuation analysis \cite{Bu1,Peng},
linguistics methods \cite{ling, Pi, Mag, Sch} {\it etc}.
Large number of models, DNA emulated, have been constructed
\cite{Her97,models}.

Although some of the studies are in contradiction with each other, the
presence of long-range
correlations and period-three oscillations in DNA sequences one
can believe statistically postulated. Now the 3 bp periodicity and the
mutual information function are widely used to find exons in newly
sequenced DNA \cite{Gr, autocor, Her97}. Long range correlations are
discussed in connection with chromosomal organization of genomes
\cite{autocor}. 

Let us notice that the main statistical investigations was performed in
the letter (single nucleotide) sequences consideration. At the same time
such elements of DNA
structure as three nucleotides (codon) protein coding, the regulatory
units such as promoters, splice sites, enhansers and silencer fairly
difficult, or with too low accuracy, can be
detected or predicted just letter analysis. So it seems to be needed to
consider regularities no only in the letter sequences but among them
groups ({\it words}) as well. As concrete example of the {\it words}
importance
one can take restrictase recognition sites as BglI:
GCC - - - - - GGC where
five
nucleotides between sequences GCC GGC may be any, or classical
Pribnow-Hilbert blocks: TTGACA- - -...- -TATAAT, where moreover, the
distance
between
the blocks can vary \cite{CAn}.

Some attempts of statistical
analysis
and classification of
short (3-8 nucleo
tid
es) sequences 
have been made \cite{words,d25,d4,Cl}. Especially it concerns three
letter 
sequences (codons), since these codons form so say amino acids (20
letters)
language. But comparing the results, one can see that
they strongly depend on an envisage object \cite{d25,Sch}. Let us
notice that triplets
analysis is also insufficient for DNA structure explanation, for example,
it tells us nothing about DNA conformational properties, interactions with
proteins and proteins-RNA complexes, equilibrium between mutation and
heredity. Though it is
reasonable to suggest that this information presents in the DNA text as
well.

If we would analyze any single genome, rather we obtain a result
appropriate for the very narrow field of investigated objects. If we would
study {\it words'} statistical properties for drastically different
organisms,
we obtain strongly
distinct results \cite{d25,d4,Cl}. For this reason we decide to pay
attention to
bacteria
genomes.
From one hand there are different kinds of bacteria that, as one can
suppose, is reflected in some distinctions of their
genomes, from the other certainly one bacteria kind
is closer to other one rather than to another species as, for example,
plants or viruses, no concerning of higher organisms. So the goal of the
present paper is
comparative analysis of bacteria genomes in {\it words} context.

In second
section we present distribution of the {\it word} frequency versus the
rank in
analogy to the Zipf analysis of natural languages \cite{Zipf}, and compare
the results with ones from linguistics DNA analysis \cite{Mag}. In Sec. 3
we pay attention to the most frequently appeared {\it words} and almost
never realized ones. 
Section 4 show the result of autocorrelation analysis.
Finally conclusions on the basis performed comparative statistical
analysis are given
in Sec. 5.

\section{Word frequency}
In analogy to Zipf analysis of natural languages \cite{Zipf} we study
distribution of {\it word's}
frequency versus rank. In order to obtain the distribution
we first rank
order the total number of occurrences of each {\it word},
and
than plot their relative values against rank. We have investigated {\it
words}
of length ($n$) from 1 to 7 nucleotides of 20 complete bacteria
genomes origin from GenBank \cite{GB}. Whose size, short
(using in this paper) and full names are
presented in the Table 1.

 The sequences length ranges from 580074 bp ({\it mgen}) to
4639221 bp ({\it ecoli}). We use the same method of sliding window 
as \cite{ling, Mag} for frequency of occurrences 
obtaining. According to this approach a window of length $n$
nucleotides/letters is taken and the set of blocks/{\it words} of size
$n$ is
obtained by shifting the window on one letter at a time. We look
for all possible {\it words} from $4^n$ for each $n \in[1,7]$.
If $n<5$ the number of all possible variants is less
300 that, generally speaking, seems to be insufficient for the statement
that the distribution obeys some law. For this reason, only for $n \in
[5,7]$ we claim that the distributions quite well approximated by
logarithmic law $$ f(r)= -\alpha 10^4 \ln r$$ where $r$ --
rank and $f$ --  frequency of occurrences of a {\it word} 
(approximation accuracy
ranges from 95.6\% ({\it hpyl} genome) to 99.6\% ({\it synecho} genome)).
Indexes $\alpha$ for each genome and $n \in[5,7]$ are presented in the
Table 2. The Figure 1 shows the distributions for {\it tpal} genome, which
has
the smallest index, {\it mgen} genome which has the largest index and
{\it aquae} genome -- an intermediate case.

Let us notice that the indexes do not depend on genome size.

Looking at the Figure 1 in \cite{Mag}, it is easy to see that the power
law
approximation no exceeds the value $10^3$ of the rank, what is only the
third part of the pointed graph, for others two parts it is obviously not
so. Other works \cite{ling} also confirm that a power
low is not the better approximation of the rank {\it word} distribution
and DNA
texts have not to believe to be written on a language in the linguistic
sense.

\section{ Most and least frequently met words}
DNA molecules consists of two strands letter sequences corresponded each
other according to the rule: versus  A letter on other strand the
letter T is situated, versus C -- G, G -- C and T -- A. It is the property
of complementarity of DNA strands.
Looking for what {\it words} are the most frequently met in the
genomes' 
texts, we obtained that
for
$n\in [2,6]$ it is polyA fragments (AAA...A $\equiv (A)_n$) (or taking
into
account sequences'
complementarity, it should be polyT fragments ($(T)_n$) as well. Let us
sign this fact as
$(A/T)_n$), where $n$
is the length of the fragment.
In the Table 3 the results of polyA/T sequences occurrence for $n \in
[2,7]$ as
well as the
genomes, whose most frequently met {\it words} is no $(A/T)_n$, are
presented.

Dominant polyA/T sequences were found in
many DNA investigations \cite{PolyA}. There are some explanations of this
phenomenon as, for instance, the fact that (A,T) relation is weaker than
(C,G), or that $(A)_n, (T)_n$  sequences have a specific
three-dimensional structure different from one of $(C)_n, (G)_n$ or other
chains, that can be necessary for nucleosomes organization \cite{Well}.
Just the
question remains why is it so only for the fragments' length less than 7 
(for the 20 complete bacteria genomes of different sizes)?

For marked by (*) genomes in the Table 3 fraction of ApT (A plus T)
nucleotides is greater than fraction of CpG. Just 4 others genomes have
 fraction of CpG greater than one of ApT.

 Below the most frequently met {\it words} for
each envisaged genome in corresponding to given in the Table 1 order are
presented:

CCTCCTC GAGGAGG AAAGGAA TTTTTCA TTTTTCT AGAAAAA CGCCAGC AAAAAAT
TTTAAAA TTTTTAA TTAAAAA TTTTTAA TCCCTGA GCCGCCG TCCTGGG TCTCCTT
TATTTTT GGCGATC GAAAGAA CGCGCGC

The most frequently met {\it word} on the level 7 (for word length equal
to 7)
among 20 genomes is TTTTTAA (2
genomes).

On 6 level:
CTCCTC GAGGAG AAGGAA TTTTTT TTTTTT AAAAAA CGCCAG AAAAAT
TTTTTT AAAAAA TTTAAA TTTAAA CCCTGA CCGCCG CTTCCT CTTCTT
ATTTTT CGATCG GAAGAA GCGCGC

That, as it can be seen, differs from the results of \cite{d4}.

For 9 from 20 investigated genomes 
({\it aero, aful, aquae, ecoli, hinf, mthe, mtub, rpxx, tpal})
the dominant {\it words} (having maximal frequency of occurrences) of
length 7 differ from ones
of length 6 
just on a single nucleotide added at the beginning or end of the {\it
word}.
8 genomes ({\it aero, aful, ecoli, mthe, mtub, pabyssi, synecho, tpal})
in the most frequent {\it words} have CpG fraction greater than
ApT.
Let us notice, mainly in the dominant {\it words} C and G letters
appear in GC/CG compositions and never we met there the fragments
(C/G)$_{k>3}$.
This can be connected with that CG (or GC) repeats in DNA in greater
degree than polyC/G fragments supply maximum contribution into free
energy of the secondary structure \cite{CAn}.

Since we look for every of $4^n$ possible words, turn out to be that no
all
of them are realized in each genome. Namely, for $n<6$ all
possible 1024 words appear at least once in every genome. For
$n=6$ three genomes have no some {\it words}: {\it hpyl} -- TCGACA
GTCGAC,
{\it mgen} -- CTCGGA CCGGCC TCGGCC GGACGC CGGCGC CCCGGC GGCCTC GCCGTC
TCCGAG CGCGCG TCGGCG GGCCGG CCTCGG GGTCGG,
{\it mjan} -- GTCGAC GCGCGC CGATCG.
For $n=7$ there are only 4 genomes containing all {\it words}:
{\it aero}, {\it bsub}, {\it ctra}, {\it tpal}. 
The number of absent {\it words} for others varies from 1 ({\it
ecoli}, {\it synecho}) to 851 ({\it mgen}).
In the Table 4 one can see the number of absent {\it words} for 
investigated
genomes.
Where from one can see that {\it words'} absence is not follow to
genome
length, as it should be for random sequences.

The rarest {\it word} (which are absent in the envisaged genomes more
often than others) for $n=6$ is GTCGAC (2 genomes),
for $n=7$ -- GCGCGCG (6 genomes), CGCGCGC GTCGACG GGCCTCG (4
genomes).

All absent and rarest {\it words} contain 
greater fraction of CpG than ApT.
At the same time neither polyC/G {\it words} nor even
(C/G)$_{k>3}$ fragments of the {\it words} are
 not absent on the level 6
or rarest on the level 7 in any from investigated genomes. In the absent
and
rarest {\it words}
quite often one can meet CG or (CG)$_k$ fragments, which as it claimed
in \cite{CAn} are more energetically profitable for secondary structure
formation.
The question remains why energetically more profitable (CG)$_k$
fragments are present in the row of the absent or rarest {\it words} but
less
energetically profitable polyG/C fragments have not been found there?

Moreover among the absent or
rarest {\it words} of length 6 there are several (complemented)
palindromes (GTCGAC, CGCGCG, GTCGAC, GCGCGC, CGATCG). That can be
understand in biological context as such palindromes 
are well known restriction enzyme cut sites, and hence
are avoided by bacteria. Thus absent and rarest {\it words} investigation
seems
to be also important part of DNA studies, since it can give us a relevant
information. 

\section{Autocorrelation analysis}

We have performed analysis of autocorrelations for the most frequently
met {\it words}. We considered length $n=3,6,7$.
For the analysis we use standard procedure of translation of
genomes' letter sequences into number representation. Namely we divide
a letter genome sequence into {\it words} of length $n$ shifting the
frame/window of length $n$
on one nucleotide for getting a new {\it word}. 
If the {\it word} on $i-th$ position is the
most frequent for an envisaged genome it is replaced by 1 in the
new representation, let us denote this fact as $x_i=1$ and $x_i=0$
otherwise. So we obtain the row of
$N-n+1$ numerical values $\{x_i\}_{i=1}^{N-n+1}$ where $N$ is genome
size.

The autocorrelation function $R(l)$ of a numerical sequence can be
written as $$R(l)=<x_i x_{i+l}>,$$
where the brackets denote average over the sites along the sequence.
Since the number of units in the chain in our case is fairly
small,
we are interested in merely the quality results, in other words,
in character of
$R(l)$ itself.

It has been obtained that for $n=3$ there are almost only correlations of
order 1 or 2. However for several genomes: {\it ecoli, mthe, mtub, tmar}
in $R(l)$ behavior oscillations of period 3 are observed.

In case $n=6$, $R(l)$ behavior acquires greater distinctions. In
the genomes
{\it synecho, pabyssi, mthe, mpneu, hinf} correlations rather are absent,
essential ones are in {\it hpyl, bsub} genomes (almost on any up to $l=50$
scale). Period 3 oscillations are present in the genomes {\it mtub, mjan,
aquae,
aful, aero}. Let us notice that for {\it mtub} genome correlations are
quite strong even for scale $l \approx 10^3$,. 
Ones are weak in
others genomes with period 3 oscillations.

As for $n=7$, in a whole in investigated genomes there is tendency of
existence of greater correlations on $l$ module 3. The most strong
correlations are found as before in {\it mtub } genome, moreover there
are existed on very large scale. In Fig. 2 one
can see $R(l)$ for this genome ($n=6$). Here also one can 
mention that {\it tpal} genome has correlations
of order 2 and 4, {\it mjan} -- 12 and 21, {\it hpyl} -- 10, 15,
21, 39, 45, {\it aero} and {\it aful} -- 3.

More detailed analysis of the genome {\it mtub} structure reveals that
period 3
oscillations are characteristic for the second and third {\it word} in
ranked
{\it words} 
distribution as well, for the forth {\it word} it is not so.
More often met {\it words} have on order higher
frequencies of occurrences for 6 from 20 envisaged genomes (for $n=7$).
Three of them ({\it hpyl, mjan, mtub}) reveal correlations.
{\it Mtub} genome has the frequency of first three {\it words} order 3 and
oscillations of period 3 are characteristic for first three {\it words} as
well. {\it
Synecho} has sharpest drop in ranked word distribution
after forth {\it word} (Fig. 3).

Such characteristic tendency of period 3 oscillations existence as in
letters as in {\it words} of different length investigations of DNA texts
can be
connected with scale invariance or
self affinity of genomes organization, in other words, DNA sequences to
all appearances posses by fractal properties. 

\section{Conclusions}
On the basis performed statistical analysis of the bacteria genomes the
main conclusions are followed.

The ranked {\it word} distributions quite well approximated
by logarithmic law.

Results obtained in absent {\it words} investigation show the considerably 
nonrandom character of DNA texts sequences and allow to reveal
biologically relevant units as restriction enzyme cut sites. That points
on importance of such kind study.

Characteristics do not depend on genome size as it has to be for random
texts.

In character of behavior of autocorrelation function in several genomes
period 3 oscillations were found. This result takes place for any
{\it word's} length from envisaged.

Short range autocorrelations are present in short ($n=3$) {\it words} and
practically absent in longer {\it words}.

Concerning autocorrelations investigation, the results obtained for
{\it mtub} genome seems to be the most interesting. Here we have strong
correlations with period 3 oscillations for any {\it word} length from
envisaged and on large scale,
that could not be detected for other genomes.

In a whole statistical analysis shows that bacteria genomes are
considerably varies from each other.
Any essential similarities for genomes of a same class (e.g. Pyrococcus:
{\it pabyssi}, {\it pyro}, Chlammydia: {\it cpneu}, {\it ctra},
Mycoplasma: {\it mpneu}, {\it mgen}) were not found.

If we want to elaborate any general scheme
of genomes classification according to statistical analysis, it will be
a fairly difficult task.
Since always there are a lot of exceptions. As for example, the {\it
words} with
CG repeats can
form as the most frequent {\it words} as never met ones. GC fragments in
the dominant {\it 
words} are more appropriate for the longer genomes ({\it ecoli, mtub, synecho})
but {\it bsub}
genome is longer than {\it synecho} but does not contain such fragments
in the dominant {\it word}.
Reasonable conclusions one can make only on the basis of as possibly
greater
set of factors. So one can
 suppose that absence of {\it words} on 6
level in
{\it hpyl} and {\it mjan} genomes (having a middle length among
investigated) is
connected with the presence of
autocorrelations of
the most frequently met {\it 
words} more than on a single scale. Strong
autocorrelations in {\it mtub} genome 
can point on
a specific structure of this genome. 
Here also one has to mention the nontypical characteristics of {\it tpal}
genome: the smallest
index in ranked {\it word} distribution, autocorrelations on 2 and 4
scales. Here the dominant
{\it word}
consists of CG repeats (that is the rarest {\it word} for
other genomes), presence of all
possible {\it words} on level 7.  All this factors allow
us to claim that this genome is the most structureless from
the investigated.

{\bf Acknowledgments.}

I would like to thank Vladimir Alenin for the useful discussion in
biological aspects of the work. This work is partially supported
by State Committee of Russian Federations for high education (grant No
97-14.3-58), St-Petersburg Government Fellowship and Soros graduate
program.

\newpage
\centerline {FIGURES}
Fig. 1 In this figure one can see the word ranked distributions
for {\it tpal, mgen, aquae} genomes in semilogarithm scale.
{\it Word} length $n=7$.

Fig. 2 In this figure one can see autocorrelation function $R(l)$ for {\it
mtub} genome in semilogarithm scale, {\it word} length $n=6$.

Fig. 3 In this figure one can see first 20 points of the word ranked
distributions for {\it hpyl, mgen, mjan, mtub, rpxx,
synecho} genomes, whose are
characterized by greater (on order) frequency of initial {\it words} and
{\it aero} genome for comparison.

\newpage
\renewcommand{\baselinestretch}{1}

\centerline {TABLES}

Table 1

\begin{center}

\begin{tabular}{|r|l|l|}
\hline
1669695& aero&	Aeropyrum pernix K1
\\
2178400 &aful&	Archaeoglobus fulgidus
\\
1551335 &aquae&	Aquifex aeolicus
\\
4214814& bsub&	Bacillus
subtilis
\\
1230230 &cpneu&	Chlamydia pneumoniae
\\
1042518& ctra&	Chlamydia trachomatis
\\
4639221 &ecoli&	Escherichia coli K-12 MG1655
\\
1830137 &hinf&	Haemophilus influenzae Rd
\\
1667867& hpyl&	Helicobacter pylori 26695
\\
 580074 &mgen&	Mycoplasma genitalium G37
\\
1664970 &mjan&	Methanococcus jannaschii
\\
 816394& mpneu&	Mycoplasma pneumoniae M129
\\
1751377& mthe&	Methanobacterium thermoautotrophicum delta H
\\
4411529 &mtub&	Mycobacterium tuberculosis
\\
1765118 &pabyssi&	Pyrococcus abyssi
\\
1738505 &pyro&	Pyrococcus horikoshii OT3
\\
1111523 &rpxx&	Rickettsia prowazekii strain Madrid E
\\
3573470 &synecho&	Synechocystis PCC6803
\\
1860725 &tmar&	Thermotoga maritima
\\
1138011 &tpal&	Treponema pallidum
\\
\hline
\end{tabular}
\end{center}

In this table size, short (using in the paper) and full names of 20
investigated bacteria genomes are presented.
\newpage

Table 2

\begin{center}

\begin{tabular}{|c|c|c|c|}
\hline
\tt name & n=5 & n=6 & n=7
\\
\hline
\tt aero & 6.12 & 1.71 & 0.49
\\
\hline
\tt aful & 5.92 &  1.62 & 0.446
\\
\hline
\tt aquae & 6.53 &  1.96 & 0.581
\\
\hline
\tt bsub & 5.87 &  1.6 & 0.452
\\
\hline
\tt cpneu & 6.27 &  1.67 & 0.47
\\
\hline
\tt ctra & 6.08 &  1.62 & 0.454
\\
\hline
\tt ecoli& 5.71 &  1.47 & 0.39
\\
\hline
\tt hinf & 7.11 &  2.04 & 0.595
\\
\hline
\tt hpyl & 7.73 &  2.26 & 0.717
\\
\hline
\tt mgen & 9.73 &  2.79m & 0.86
\\
\hline
\tt mjan & 10.04m &  2.78 & 0.857
\\
\hline
\tt mpneu & 6.94 &  1.97 & 0.572
\\
\hline
\tt mthe & 6.43 &  1.7 & 0.468
\\
\hline
\tt mtub & 8.6 &  2.44 & 0.725
\\
\hline
\tt pabyssi & 6.25 &  1.62 & 0.438
\\
\hline
\tt pyro & 7.39 &  1.87 & 0.497
\\
\hline
\tt rpxx & 9.21 &  2.65 & 0.84
\\
\hline
\tt synecho & 6.14 &  1.62 & 0.448
\\
\hline
\tt tmar & 6.52 &  1.84 & 0.536
\\
\hline
\tt tpal & 4.49m &  1.25m & 0.356m
\\
\hline
\end{tabular}
\end{center}

In this table indexes $\alpha$ for each envisaged genome and {\it
word's} length $n \in[5,7]$ are presented.
\newpage

Table 3

\begin{center}
\begin{tabular}{|c|c|l|}
\hline
2 &14/20 &aero, eco
li, mthe$^{(*)}$, mtub, tmar$^{(*)}$, tpal
\\
3 &12/20 & + aful, pabyssi
\\
4 &11/20 & + pyro
\\
5 &10/20 & + aquae
\\
6 &5/20 & + hinf, mjan, mpneu, rpxx, synecho
\\
7 &0& all
\\
\hline
\end{tabular}
\end{center}

In this table the results of polyA/T sequences occurrence for $n \in
[2,7]$ as well as the
genomes, whose most frequently met {\it words} is no $(A/T)_n$, are
presented.

The first column of the table shows the level number (length of
the {\it word}); the second column -- ratio of the number of the genomes,
whose
have dominant polyA/T sequence, to total number of genomes; the third
column
-- the names of the genomes, whose most frequently met {\it words} is no
$(A/T)_n$.

(We present in this table the genomes whose most frequently met {\it
words} are
another than polyA/T because for $n=2$ the number of such genomes is less,
therefore, since if a genome has no polyA/T sequence as dominant on
the second level ($n=2$) it has no one as dominant on any higher level. For
$n=3$ we only add to the genomes from previous level such ones, who has
dominant polyA/T on the second level and has no it on the third and so
on up to the level 7.)
 
\newpage
Table 4

\begin{center}
\begin{tabular}{|c|c|c|c|c|c|c|c|c|c|}
\hline
aero &aful &aquae &bsub &cpneu& ctra&ecoli &hinf &hpyl & mgen
\\
0  &  4 &   4  &   0 &   2 &    0  &  1 &    11 &  192 &  851
\\
\hline
mjan& mpneu& mthe& mtub& pabyssi& pyro& rpxx& synecho& tmar& tpal
\\
318 & 7&     5&    3&    3&       4&    71 &  1  &     2 &   0
\\
\hline
\end{tabular}
\end{center}

In this table one can see the number of absent {\it words} for envisaged
genomes.


\newpage
\begin{thebibliography}{50}
\vspace{-0,2cm}

\bibitem{GB} URL: http://www.ncbi.nlm.nih.gov

\bibitem{JW} J.T.L. Wang, S. Rozen, B.A. Shapiro, D. Shasha, Z. Wang, M.
Yin, J. Comput. Biol. 6 (1999) 209. 

W.R. Pearson, D.J. Lipman, Proc. Nat. Acad. Sci. USA 85 (1988) 2444.

T.F. Smith, M.D. Warterman, J. Mol. Biol. 147 (1981) 195.

\bibitem{autocor} H. Herzel, E.N. Trifonov, O. Weiss, I. Grosse, Physica A
 249 (1998) 449. 
\bibitem{MA} M. de Sousa Vieira, Phys. Rev. E 60 (1999) 5932. 
\bibitem{Bor} B.Borstnik, D. Pumpernik, D. Lukman, Europhys. Lett.
23 (1993) 389. 
\bibitem{Voss} R. Voss {\it et. al.}, Phys. Rev. Lett. 68 (1992) 3805;

A.A. Tsonis, J.B. Elsner, P.A. Tsonis, J. Theor. Biol. 151 (1991) 323;

V.R. Chechetkin, A. Yu. Turygin, J. Phys. A 27 (1994) 4875. 

\bibitem{Bu1}S.V.
Buldyrev {\it et. al.}, Phys. Rev. E 51 (1995) 5084.

\bibitem{Bu2} C.-K. Peng, S.V. Buldyrev, A.L. Goldberger, S. Havlin, F.
Sclortino, M. Simonis, H.E. Stanley, Nature 356 (1992) 168.

\bibitem{wawlet}A.A. Tsonis, P. Kumar, J.B. Elsner, P.A. Tsonis, Phys. Rev. E
53 (1996) 1828;

A. Areneodo, E. Bacry, P.V. Graves, J.F. Muzy, Phys. Rev. Lett. 74
(1995) 3293 ;

A. Areneodo, Y. d'Aubenton-Carafa, E. Bacry, P.V. Graves, J.F. Muzy, C.
Thermes, Physica D 96 (1996) 291. 

\bibitem{H94} H. Herzel, W. Ebeling, A. O. Schmitt, Phys. Rev. E 50 (1994) 5061.

\bibitem{Rowe}G.W. Rowe,  J. Theor. Biol. 101 (1983) 151.
\bibitem{Gl} L.L. Gatlin,  J. Theor. Biol. 10 (1966) 281.
\bibitem{Lu} X. Lu, Z. Sun, H. Chen, Y. Li, Phys. Rev. E 58 (1998) 3578.
\bibitem{Yu}Z-G. Yu, G-Y. Chen, appear in Communication in theoretical physics
(2000)
\bibitem{rw} S. Karlin, V. Brendel, Science 259 (1993) 677.
\bibitem{Her97} H. Herzel, I. Grosse, Phys. Rev. E 55 (1997) 800.
\bibitem{CSF}  H. Herzel, A.O. Schmitt, W. Ebeling, Chaos, Solitons, Fractals 
4 (1993) 97.
\bibitem{Peng} C.-K. Peng, S.V. Buldyrev, S. Havlin, M. Simonis, H.E. Stanley,
A.L. Goldberger, Phys. Rev. E 49 (1994) 1685.
\bibitem{ling} A.A. Tsonis, J.B. Elsner, P.A. Tsonis, J. Theor. Biol.
184 (1997) 25.
\bibitem{Pi} S. Pietrokovski, J. Biotechnol. 35 (1994) 257.
\bibitem{Mag}R.N. Mantegna, S.V. Buldyrev, A.L. Goldberger, S. Havlin, M. Simonis,
C.-K. Peng, H.E. Stanley, Phys. Rev. Lett. 73 (1994) 3169.
\bibitem{Sch} A.O. Schmitt, W. Ebeling, H. Herzel, BioSystems 37 (1996)
199  
\bibitem{models} 
W. Li, Europhys. Lett. 10 (1989) 395;

W. Li, Phys. Rev. A  43 (1991) 5240;

W. Li, Computers Chem. 21 (1997) 257;

W. Li, T. G. Marr, K. Kaneko, Physica D 75 (1994) 392;

G. A. Churchill, Bull. Math. Biol. 51 (1989) 79;

G. A. Churchill, Computers and Chemistry 16 (1992) 107;

D.G. Arques, C. J. Michel, K. Orieux, J. Theor. Biol. 161 (1993) 329;

S. Karlin, V. Brendel, Science 259 (1993) 677.

\bibitem{Gr} I. Grosse, H. Herzel, S.V. Buldyrev, H.E. Stanley, Phys. Rev.
E 61 (2000) 
\bibitem{CAn} A.A.Alexandrov, N.N.Alexandrov, M.Yu. Borodovsky et al., 
Computer analysis of gene texts (Nauka, Moscow, 1990).

\bibitem{words} G.J. Phillips, J. Arnold, R. Ivarie, Nucl. Acids Res. 15
 (1987) 2611; {\it ibid} 15 (1987) 2627;

R. Grantham, C. Gautier, M. Gouy, M.
Jacobzone R. Mercier, Nucl. Acids Res. 9 (1981) 243;

E. N. Trifonov V. Brendel, GNOMIC: a dictionary of the genetic code
(Balaban Publishing, Philadelphia, 1986);

T. Ikemura, J. Mol. Biol. 146 (1981) 1;

S. Ohno, Proc. Natl. Acad. Sci. U.S.A. 85 (1988) 4378.

\bibitem{d25} S.Tavere, B. Song, Bull. Math. Biol. 51 (1989) 95;

T. Yomo, S. Ohno, Proc. Natl. Acad. Sci. U.S.A. 86 (1989) 8452.

\bibitem{d4} S. Karlin, G. Ghandour, Proc. Natl. Acad. Sci. U.S.A.
82 (1985) 6186; {\it ibid} 82 (1985) 5800;
\bibitem{Cl} J-M. Claverie, L. Bougueleret, Nucl. Acids Res. 14
 (1985) 179.

\bibitem{Zipf} G.K. Zipf, The psycho-biology of language
(The M.I.T. Press, Massachusetts, 1965).

\bibitem{PolyA} R. Nussinov, Nucl. Acids Res. 8 (1980) 4545;

R. Nussinov, J. Theor. Biol. 85 (1980) 285;

J. Moreau, et al. Nature 295 (1982) 260.

\bibitem{Well} R.D. Wells et al., CRC Critical Reviews in Biochemistry
305 (1977).
\end{thebibliography}
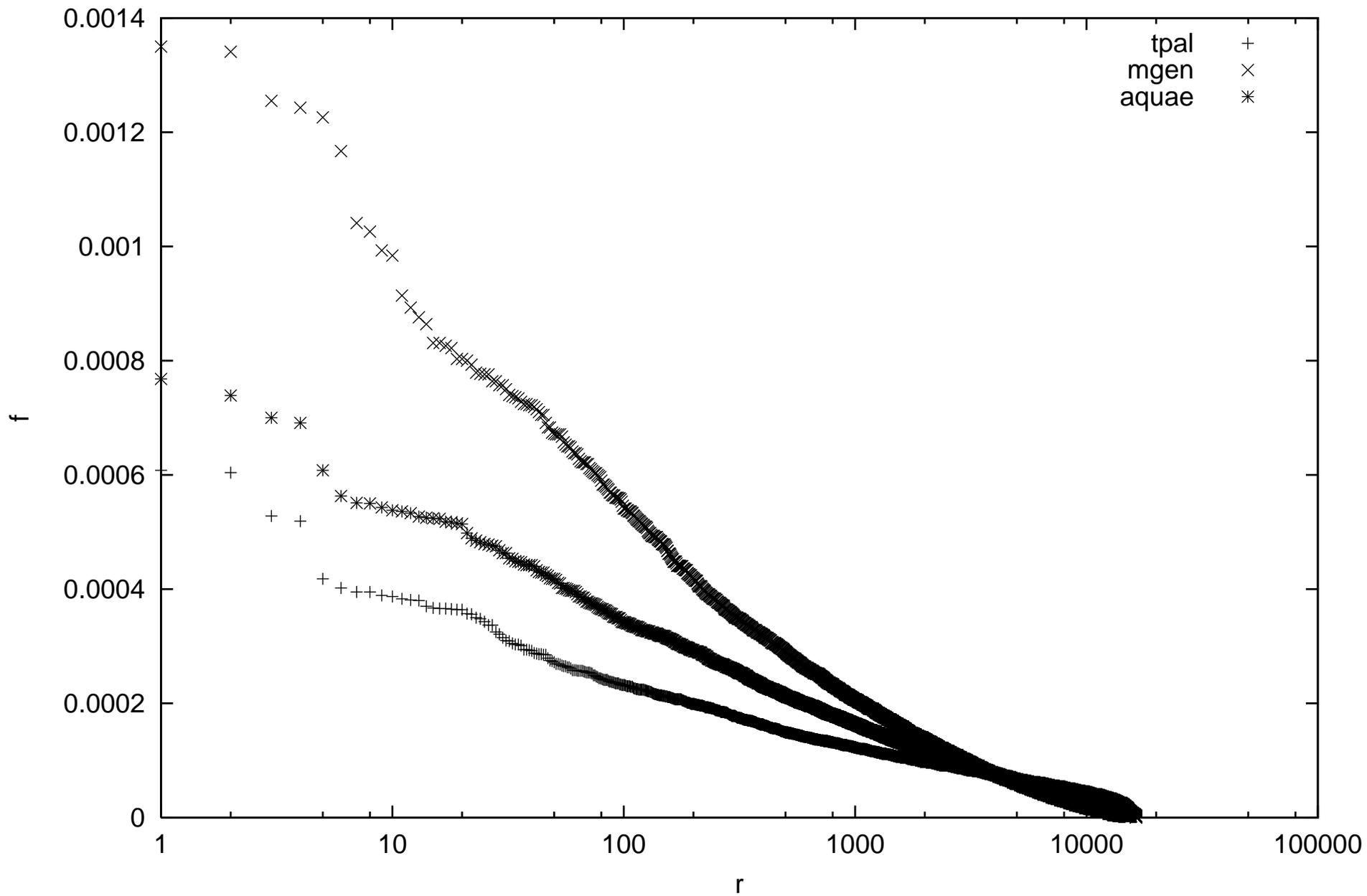
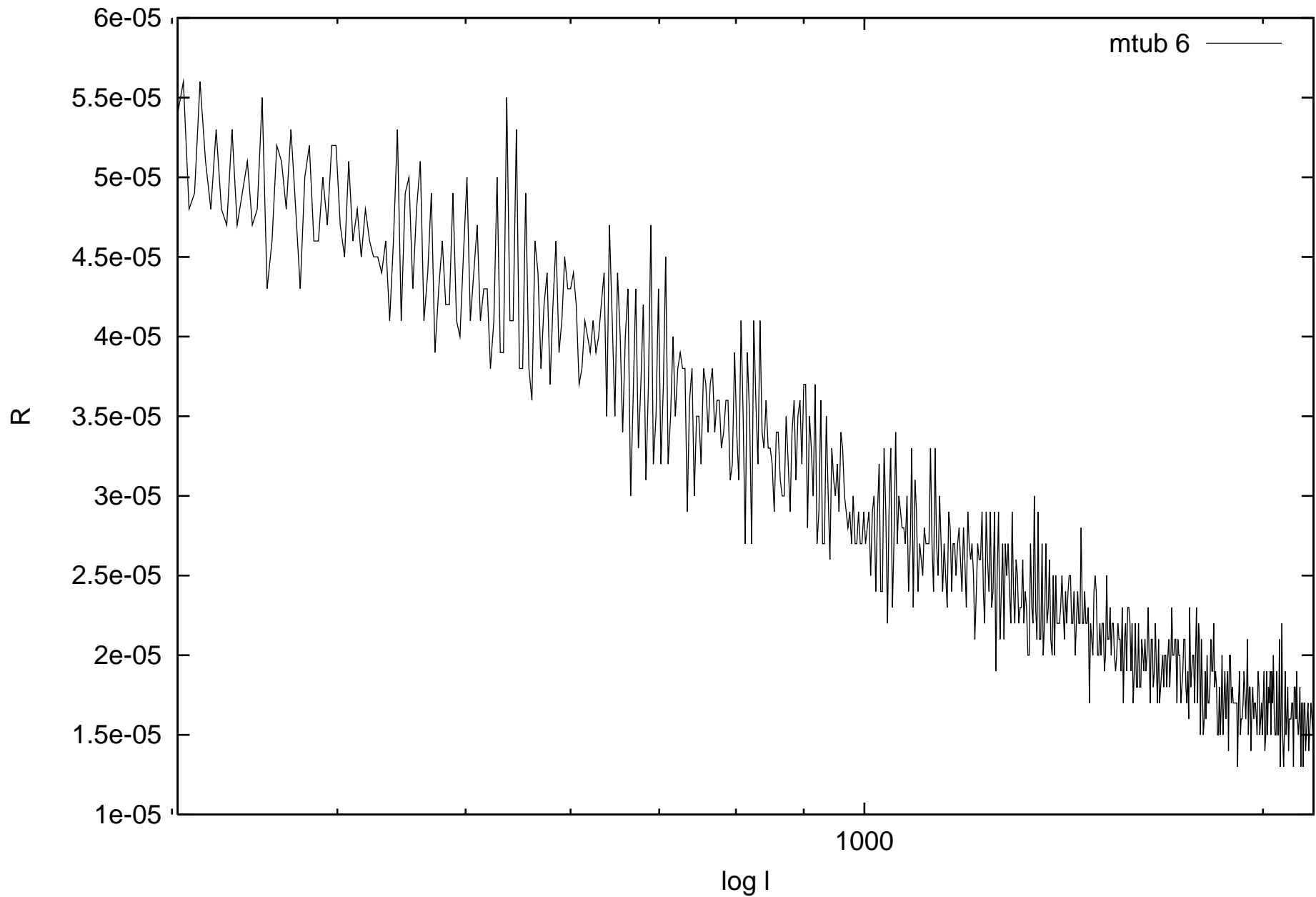
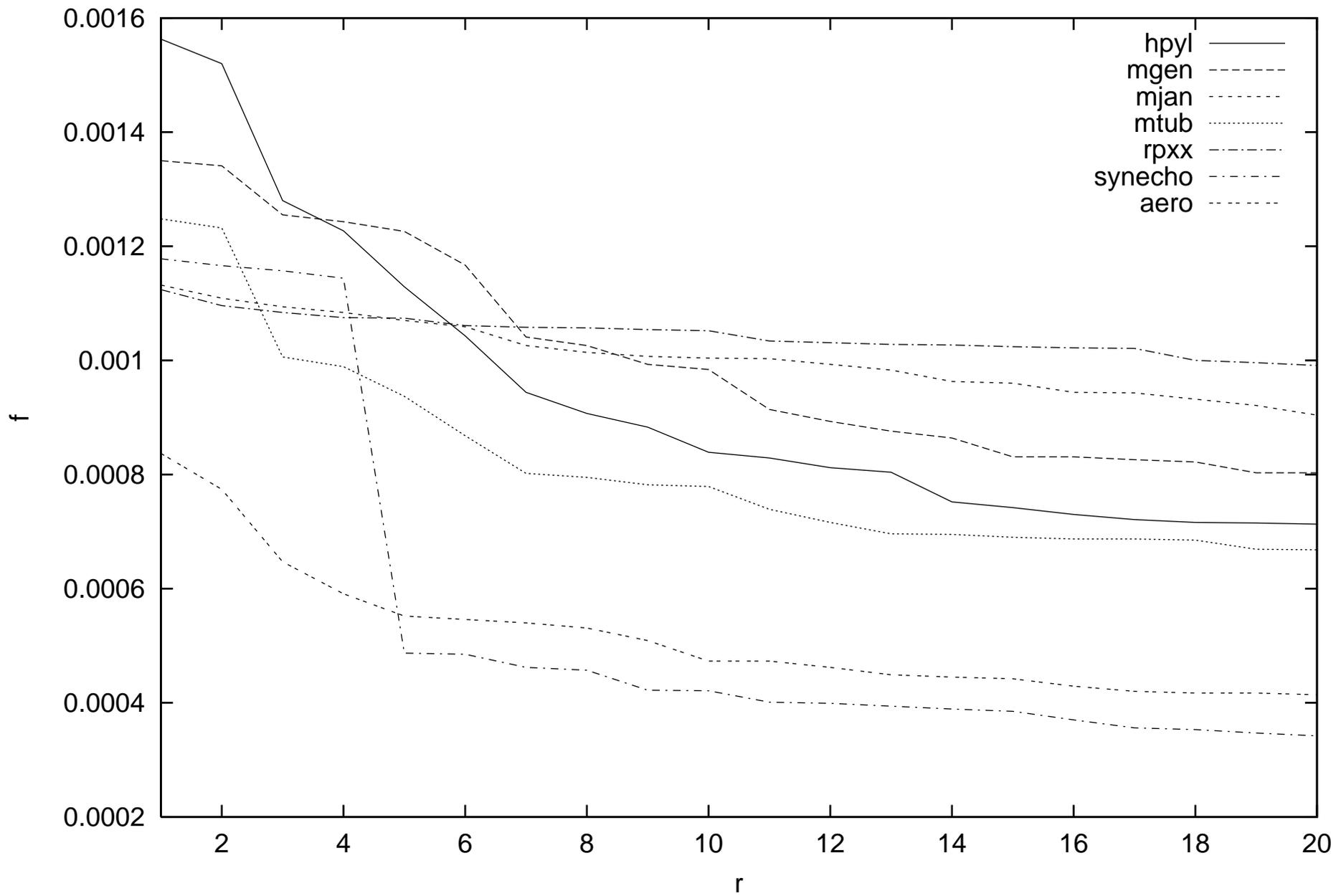
\end{document}